\documentclass[preprint,final]{aastex}
\shorttitle{NGC 6791 Variability Study}
\shortauthors{Hartman et al.}
\begin{document}

\title{Pushing the Limits of Ground-Based Photometric Precision -- Sub-Millimagnitude Time-Series Photometry of the Open Cluster NGC 6791}

\author{J.~D.~Hartman, K.~Z.~Stanek, B.~S.~Gaudi, M.~J.~Holman, B.~A.~McLeod}
\affil{Harvard-Smithsonian Center for Astrophysics, 60 Garden St.,
Cambridge, MA~02138}
\email{jhartman, kstanek, sgaudi, mholman, bmcleod@cfa.harvard.edu}

\begin{abstract}

We present the results from a three night, time-series study of the
open cluster NGC 6791 using the Megacam wide-field mosaic CCD camera
on the 6.5m MMT\footnote{Observations reported here were obtained at
the MMT Observatory, a joint facility of the Smithsonian Institution
and the University of Arizona.} telescope. The aim of this study was to
demonstrate the ability to obtain very high precision photometry for a
large number of stars. We achieved better than 1\% precision for more
than 8000 stars with $14.3<R<20.1$ and sub-millimagnitude (as low as 360 $\mu$mag)
precision for over 300 stars with $14.6<R<16.3$ in the field of this cluster. We also
discovered 8 new variable stars, including a possible $\delta$-Scuti
variable with an amplitude of 2\%, 6 likely W UMa contact binaries,
and a possible RS CVn star, and we identified 5 suspected
low-amplitude variables, including one star with an amplitude of 3
mmag. We comment on the implications of this study for a ground-based
survey for transiting planets as small as Neptune.

\end{abstract} 

\section{Introduction}

Since the advent of high quality CCDs there has been a great deal of
effort to obtain high-precision (millimagnitude, or less than 1\%)
time-series, optical and infrared photometry for large numbers of
stars. Photometry at the millimagnitude level of precision is
considered a prerequisite in searches for transiting Jupiter-sized
planets around solar-type stars. As a result, the number of groups
that have achieved this level of precision is too many to list here
(see for example \citealt{horne03} for a list of transiting planet
searches). The search for transiting planets has not been in vain; at
the time of writing there are 7 known transiting planets, 6 of which
were first identified photometrically (see e.g. \citealt{udalski04}
and \citealt{konacki04}). Millimagnitude photometry has also
contributed to the study of stars near the hydrogen fusion limit
\citep{pont05} and enabled the study of stellar variability with
amplitudes of a few percent (e.g. \citealt{bruntt03}). With the number
of transit surveys growing steadily, millimagnitude photometry has
become routine. Pushing below $0.1$\%, however, has proven difficult.

To our knowledge, \citet{gilliland93} holds the record for the most
precise (per exposure) ground-based photometry that has been
reported. They achieved a precision as good as 250 $\mu$mag per
exposure for a group of 12 stars in M67 that they monitored for
solar-like oscillations. Prior to that, \citet{gilliland92} achieved a
precision of 750 $\mu$mag per exposure. In both of these cases only a
handful of bright isolated stars were monitored. Since these projects
were aimed at searching for short time-scale, solar-like oscillations,
the authors applied high-pass filters to their light curves, thereby
removing any long time-scale systematic trends together with any long
time-scale variability. Ground-based, sub-millimagnitude per exposure
photometry has also been obtained for individual bright objects
(e.g. \citealt{jha00} who obtained an RMS of 800 $\mu$mag for the
transiting system HD209458 using a photometer; also see \citealt{kurtz05}).

Despite the difficulties in performing sub-millimagnitude photometry
from the ground for large numbers of stars, the possible science
rewards are compelling. Improving the precision of transit surveys by
a factor of ten would allow for the detection of $0.1$\% transits due
to Neptune-sized planets orbiting solar-type stars. It would also
allow the exploration of a new regime of stellar variability.

As discussed in the next section, the Megacam instrument on the MMT
telescope is an ideal setup for achieving sub-millimagnitude
photometry from the ground. Motivated by the possibility of opening a
new regime to ground-based, time-series campaigns, we set out to
demonstrate photometry for a large number of stars with a per-exposure
precision as good as a few parts in 10,000 by conducting a short
time-series study of the open cluster NGC 6791 using MMT/Megacam.

In the following section we describe our observations. We follow with a discussion of our data reduction steps in \S 3; in \S 4 we describe the photometric precision we have achieved; in \S 5 we present new variable stars that we have found in this field; and we finish with a discussion of our results, including the possibility of a search for transiting Neptune-sized planets in \S 6.

\section{Observations}

The data for this project were obtained on the nights of October 4th,
9th and 20th of 2004 using the Megacam CCD mosaic \citep{mcleod00}
mounted on the MMT 6.5m telescope. The Megacam instrument is a
$24\arcmin$ x $24\arcmin$ mosaic consisting of 36 2k$\times$4k,
thinned, backside-illuminated CCDs that are each read out by two
amplifiers. The mosaic has a pixel scale of $0.08\arcsec$ which allows
for a well sampled point-spread-function (PSF) even under the best
seeing conditions. The result is that in $1\arcsec$ seeing one can
collect as many as $2 \times 10^{7}$ photons from a single star prior
to saturation, setting the photon limit for the precision in a single
exposure at 0.25 mmag.

For this study we chose to observe the open cluster NGC 6791. This
cluster has been studied extensively for variability by the PISCES
project (\citealt{mochejs02,mochejs05}, additional
variability surveys of this cluster include those by
\citealt{kaluzny93}, \citealt{rucinski96}, \citealt{mochejs03}, and
\citealt{bruntt03}). As noted in \citealt{mochejs05}, the cluster is
populous \citep{kaluzny92}, old ($\tau=8$Gyr), metal rich
([Fe/H]=$+0.4$), and located at a distance modulus of
(m-M)$_{V}=13.42$ \citep{chaboyer99}.

\begin{figure}[p]
\epsscale{1}
\plotone{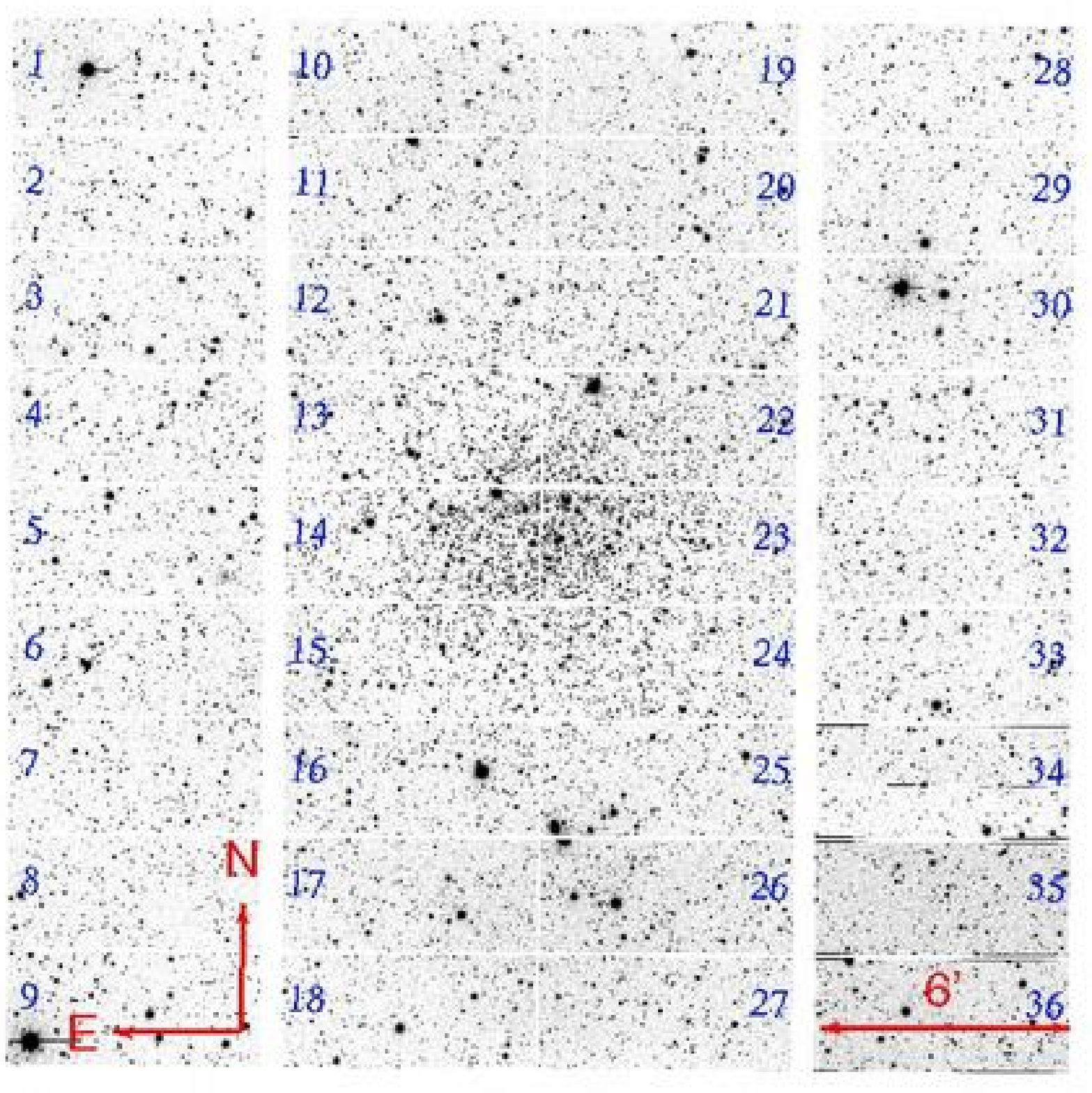}
\caption{Megacam mosaic image of the open cluster NGC 6791. Chip labels are for reference in the text.}
\label{fov}
\end{figure}

We obtained 71 exposures centered on the cluster
$(\alpha,\delta)=(19^{\rm h}20^{\rm m}53\fs0,
+37\arcdeg46\arcmin30\farcs0)$ (J2000.0) using a Sloan-$r^{\prime}$
filter. Of these exposures, 20 were obtained on Oct. 04, 2004 with a
two minute exposure time, 17 on Oct. 09, 2004 with a two minute
exposure time, and 36 on Oct. 20, 2004 with a one minute exposure
time. The data on the first two nights were obtained with a gain
setting of 10 e$^{-}$/ADU, after noting the possibility of
nonlinearity in pixels with more than $2\times 10^{5}$~e$^{-}$ we
switched to a gain of 3.5 e$^{-}$/ADU for the last night. In both
cases we were not limited by the A/D converter. All images were
read-out using 2x2 binning (yielding a pixel scale $0.16\arcsec$), but
this did not limit the number of electrons in the detector. For
reference, we present a Megacam mosaic image of the field in
Fig.~\ref{fov}.

For the first night the seeing was highly variable, ranging from
$1\arcsec$ to $3\arcsec$. On the second and third nights the seeing
was relatively stable, but not exceptional, and ranged from $1\arcsec$
to as high as $2\arcsec$ in a handful of images. The poor conditions
on the first night make the data unusable for precision photometry
using our reduction techniques, though we include data from this night
in the light curves presented in \S 5.

\section{Data Reduction}

The preliminary CCD reductions, including overscan, zero level
correction, and flat-fielding were performed using the standard
routines in the IRAF MSCRED package\footnote{IRAF is distributed by
the National Optical Astronomy Observatories, which is operated by the
Association of Universities for Research in Astronomy, Inc., under
agreement with the National Science Foundation.}. For each night we
constructed a master twilight flat-field from 5, 19, and 5 individual
twilight flat-field exposures, respectively.

To obtain photometry we used the image subtraction methods due to
Alard \& Lupton (1998; see also Alard 2000) as implemented in the ISIS
2.1 package\footnote{The ISIS package is available from C.~Alard's Web
site at http://www2.iap.fr/users/alard/package.html.}. The procedure
we followed is similar to that described in e.g. \citet{hartman04};
here we only highlight differences from the procedure discussed there. The basic scheme is to match the PSF and background of a reference image to another image, subtract them, and perform photometry on the subtracted image. The photometry routine that comes with the ISIS package convolves a PSF determined empirically on the reference image with the convolution kernel used to match the images, and then performs fixed-position, PSF fitting photometry on the subtracted image.

We performed subtraction independently for 33 of the 36 CCDs in the
mosaic (the three chips labelled 34-36 in
Fig.~\ref{fov} had artifacts that rendered them unusable at the time), dividing
each CCD into two independent sub-regions. We created saturation masks
for each image, so that pixels above 60,000 ADU would not be used in
the photometry extraction routines. Prior to registering the images we
binned them 3x3. This reduced the full width at half maximum (FWHM) of
the PSF to 2-3 pixels over the range of seeings that were encountered,
and thus allowed us to use the existing ISIS routines without
substantial modification. Reference images for each chip were created
from the best seeing images on the third night. 

Because one only measures differential flux with image subtraction, it
is necessary to obtain base fluxes for the stars via another technique
if one wishes to obtain light curves in magnitudes. We obtained these
fluxes by performing PSF photometry on the reference images for each
chip using DAOPHOT/ALLSTAR (Stetson 1987, 1992). To ensure that the
fluxes that we measured in ISIS are on the same scale as those
measured with DAOPHOT/ALLSTAR, we performed an aperture correction by
measuring the flux on the reference images using aperture photometry
with a radius equal to that used in ISIS, and adjusting the PSF
photometry to remove any systematic differences from the aperture
photometry. The corrections were typically less than 0.1mag (in
absolute value), and all had an RMS uncertainty less than 10 mmag. As
a result, any systematic error in the amplitudes should be less than
$1$\% of the stated amplitude.

We proceeded to obtain photometry for 27,885 stars on 33 CCDs with
$13.9<R<24.0$. We calibrated the photometry to the R-band using data
provided to us from PISCES (B.~Mochejska, private communication,
2005). Having obtained the light curves, we performed two cleaning
steps. The first step was to remove any systematic changes in the
zero-points of the light curves. We did this by solving for the
corrections to the zero-points that minimize
\begin{equation}
\sum_{i,j}\frac{(m_{ij}+\Delta z_{i}-\bar{m_{j}})^{2}}{\sigma_{ij}^{2}},
\end{equation}
where $m_{ij}$ is the magnitude of the $j$th star at time $i$,
$\bar{m_{j}}$ is the average magnitude of the $j$th star,
$\sigma_{ij}$ is the uncertainty in the magnitude of the $j$th star at
time $i$ and is calculated assuming photon noise and the gain listed
in the Megacam Observers Manual, and $\Delta z_{i}$ is the correction
to the zero-point at time $i$. The zero-point corrections were all
less than 1 mmag and hence only affected the precision of the
brightest stars. In the second step we rejected observations, for each
chip, for which a substantial percentage of stars on the chip (more
than 4.5\%) showed a greater than $3\sigma$ deviation from their
mean. This removed 3 or 4 of the 36 observations on the third night
for most of the chips. The rejected observations were among the
poorest seeing images for the night. For the second night we used a
less stringent criterion of 19\% to remove 1 or 2 of the 17
observations for most of the chips. We required a less stringent
criterion because a greater fraction of the stars for which we extracted
photometry were saturated in the longer exposures of the second night.

Even with the aperture correction it is still possible for there to be
a systematic error in the amplitudes of variations. This would happen,
for example, if there was a systematic error in the ISIS photometry
that might result from errors in the subtraction process. To ensure
that we are not underestimating the amplitudes of our light curves,
and hence overestimating our precision, we extracted photometry for a
handful of simulated variable stars.

\begin{figure}[p]
\epsscale{1}
\plotone{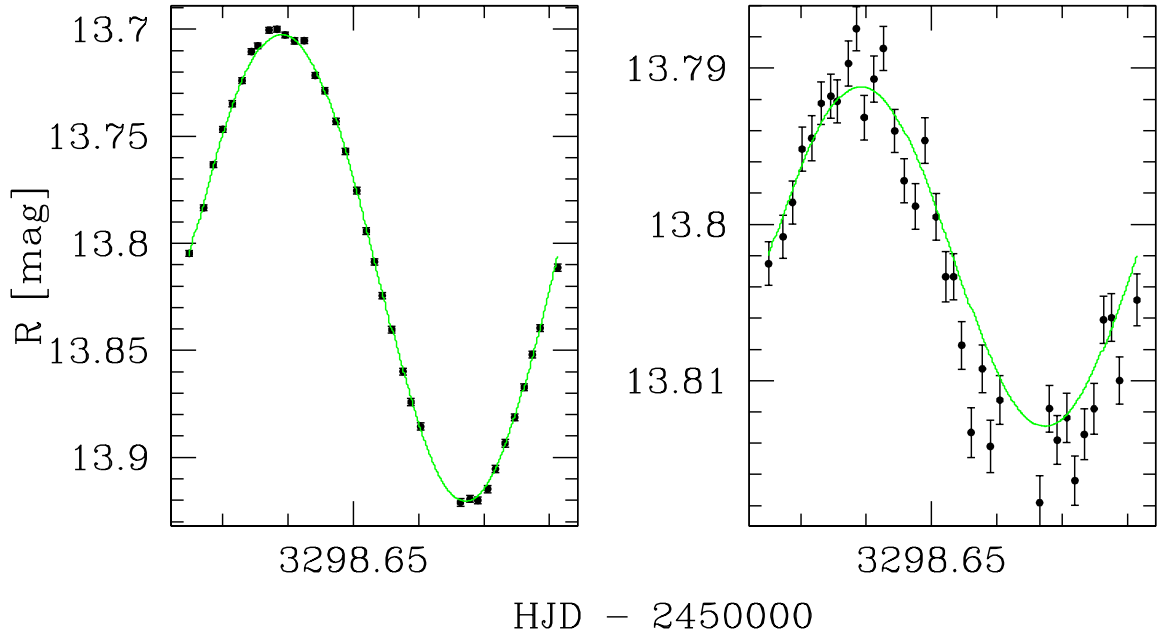}
\caption{Light curves of simulated variable stars. Points were extracted using image subtraction, the solid line shows the input variable star signal. The noise in the light curve comes from the base star that is multiplied by the signal, as well as from errors in the simulation process. At left is a 10\% semi-amplitude variation, at right is a 1\% semi-amplitude variation. Note that the amplitude of the measured light curve is not less than that of the input signal.}
\label{fakelcs}
\end{figure}

To add the simulated variable stars we first identified a bright
isolated star on one of the images and extracted a small box around
the star in every image. We then measured and subtracted the sky from
the box, multiplied the box by a scaling factor and added the result to
another location on the image. In this way we simulated two variables
stars with semi-amplitude flux-variations of 10\%, and 1\%. We present
the resulting light curves in Fig.~\ref{fakelcs}. The purpose of this
procedure was to test for systematic errors in the amplitudes, we
stress that the overall noise in the light curves is not
representative of noise expected for stars of this brightness as extra
noise is introduced in the sky subtraction process. As is apparent
from Fig.~\ref{fakelcs}, the light curves are in good agreement with
the simulated signal.

\section{Photometric Precision}

In Fig.~\ref{lcstat_separate} we plot the RMS of each light curve
versus the average magnitude for that light curve. We plot each night
separately, for both the entire mosaic and the best individual chip
(labeled 21 in Fig.~\ref{fov}).

For reference we also plot the $6.5\sigma$ detection limits for Jupiter and Neptune sized transiting planets. These lines are defined by eq.~[3] in \citet{mochejs02}
\begin{equation}
{\rm RMS} = \sqrt{N}\Delta R/\sigma,
\end{equation}
where $N$ is the number of observations in transit, $\Delta R$ is the amplitude of the transit, and $\sigma$ is set to $6.5$. To calculate $N$ we use the length of a transit as given by eq.~[1] of \citet{gilliland00}
\begin{equation}
\tau = 1.412M_{*}^{-1/3}R_{*}P^{1/3},
\end{equation}
where $\tau$ and $P$ are in days, and the stellar mass, $M_{*}$, and radius, $R_{*}$, are in solar units. To obtain $M_{*}$ and $R_{*}$ as functions of magnitude we generated isochrones from \citet{girardi00} using the parameters for the cluster listed in \S 2. The lines were then calculated for a $P=3.5$\/day period planet assuming one observes 3 full transits with two minute integrations taken every 3 minutes for comparison with night 2, and one minute integrations taken every 2 minutes for comparison with night 3. The implications of these lines are discussed in \S 6.

There are a total of 378 stars that have RMS $<$ 1 mmag on the second
night, and 9661 with RMS $<$ 10 mmag. For the third night, with the
shorter exposure times, we find 365 stars with RMS $<$ 1 mmag and 8132
with RMS $<$ 10 mmag. When the two nights are combined we find only 65
stars with RMS $<$ 1 mmag and 7732 with RMS $<$ 10 mmag. The drastic
reduction in the number of stars with RMS $<$ 1 mmag is not unexpected
as the differing exposure times between the two datasets results in a different
saturation level.

From Fig.~\ref{lcstat_separate} it is clear that the observed RMS values
are consistent with photon statistics for all but the brightest
magnitudes. For stars brighter than $R=16$ mag there appears to be an
additional source of error contributing to the RMS. To determine this constant error for each chip on the second and third nights we estimated
the RMS in magnitudes of the jth light curve as:
\begin{equation}
\sigma_{j} = \sqrt{\sum_{i=1}^{n}\frac{(2.5\log_{10}({\rm e}))^2(F_{ji}+s_{i})}{g_{eff}\cdot n\cdot F_{ji}^{2}}+c^{2}},
\end{equation}
where $n$ is the number of images, $F_{ji}=10^{2(z_{i}-m_{ji})/5}$ is the flux in ADU of the $j$th light curve on the $i$th image, $z_{i}$ is the zero-point of the $i$th image, $m_{ji}$ is
the magnitude of the $j$th light curve on the $i$th image,
$s_{i}$ is the effective sky flux of the $i$th image, $g_{eff}$ is the effective gain of the
chip, and $c$ is the constant error term for the chip. When performing PSF fitting the above equation is applicable except that the effective gain is less than the actual gain, with the exact factor depending on the shape of the PSF and the size of the region one uses to fit the PSF (see \citealt{kjeldsen92} eq.~[37] for the case of a Gaussian PSF). We find values for the effective gain that are typically less than the actual gain of the CCD by a factor of $\sim 1.7$.

We find that on the second night the constant error term ranges from
0.56 mmag to 2.4 mmag with an average value (over stars) of 1.2 mmag,
while for the third night the constant error term ranges from 0.42
mmag to 1.6 mmag with an average value of just below 1 mmag. We also
note that stars faint enough for the errors to be dominated by photon
statistics have a lower RMS in the second night compared to the third
by a factor that is consistent with the longer exposure time for the
second night.

When the data for the second and third nights are combined, the RMS is
not increased for stars that are below saturation in both nights ($R <
15.5$). This implies that there is no substantial systematic offset
between the nights and suggests that one may be able to consistently
achieve this level of precision for a longer time series campaign.

There are a number of possible sources for the observed constant error
term in our photometry. The relative error in the photometry (in
magnitudes) due to Poisson noise in the flat-field is
\citep{kjeldsen92}:
\begin{equation}
c_{ff} = \frac{2.5\log_{10}({\rm e})}{N_{ff}^{1/2}}
\end{equation}
where $N_{ff}=\Omega_{eff}e_{ff}$, $\Omega_{eff}$ is the effective
area of the PSF in pixels, and $e_{ff}$ is the total number of
electrons in the flat-field in one pixel. For the third night the
combined flat-field has $e_{ff} = 3.3 \times 10^{5}$ electrons/2x2
pixel, and the FWHM ranged from 6 to 12 2x2 pixels. Thus the expected
constant error term due to flat-fielding lies below 0.2 mmag, well
below the measured constant error terms for that night. This
calculation assumes that the flat-fielding error is dominated by shot
noise, the actual error may be larger if there are other systematic
errors in the flat-field. The effect of this error is reduced when the
pointing is stable between images. We note that because the PSF is so
well-sampled on the Megacam CCDs we do not expect intra-pixel
variations in the quantum efficiency to make a significant
contribution to the error.

Atmospheric scintillation also adds an effective constant error term
to the photometry. This error can be estimated from \citet{young67} as
\begin{equation}
c_{scint}=0.1d^{-2/3}X^{1.75}\exp(-h/8000)(2t_{exp})^{-1/2}
\end{equation}
where we use $c_{scint}$ for the constant error term due to
scintillation, $d$ is the telescope diameter in cm, $X$ is the
airmass, $h$ is the observatory altitude in m, and $t_{exp}$ is the
exposure time in s. The leading coefficient is rather approximate (we
multiply by $2.5\log_{10}({\rm e})$ to convert to magnitudes), as
scintillation can change by a factor of 2 in a few minutes
(e.g. \citealt{young93}). For the second night, our observations were
60 seconds long, with the airmass ranging from 1.12 to 1.37. The MMT
is located at an altitude of 2606 m, and has a diameter of 650
cm. Therefore we expect a constant error term of less than about 0.15
mmag due to atmospheric scintillation, and thus the total constant
error term should be less than 0.25 mmag.

Because one obtains the PSF only once on the reference, the error in
the PSF will not contribute to the errors in differential
photometry. Instead, errors in the kernel that is used to convolve the
PSF effectively add a constant error term to the photometry. It is
difficult to estimate a priori exactly what that error should be, and
we suspect that the constant error term that we have measured is due
to this effect. To test this hypothesis we have also performed simple
unit-weight aperture photometry on the subtracted images, the results
for a single chip on the third night are shown in
Fig~\ref{apcompare}. To correctly scale the fluxes we divided by the
integral of the PSF over the aperture radius. We varied the aperture
radius to optimize the precision at the bright end while providing
correct amplitudes for the simulated variable stars mentioned in the
previous section. The aperture photometry light curves were then put
through the cleaning procedures described in the previous section to
provide a fair comparison with the optimal PSF light curves. As
expected, unit-weight aperture photometry performs worse than PSF
fitting for faint stars as it is subject to a greater degree of sky
background, however for the brightest stars aperture photometry
outperforms PSF fitting, and appears to show no evidence of a constant error term. This effect is well-known when not using image
subtraction, and confirms our suspicions that the constant error term
arises from uncertainties in the kernel propagated through PSF
fitting. Using aperture photometry on the subtracted images we achieve
a precision as good as 360 $\mu$mag per exposure.

\section{Variable Stars}

\begin{figure}[p]
\epsscale{1}
\plotone{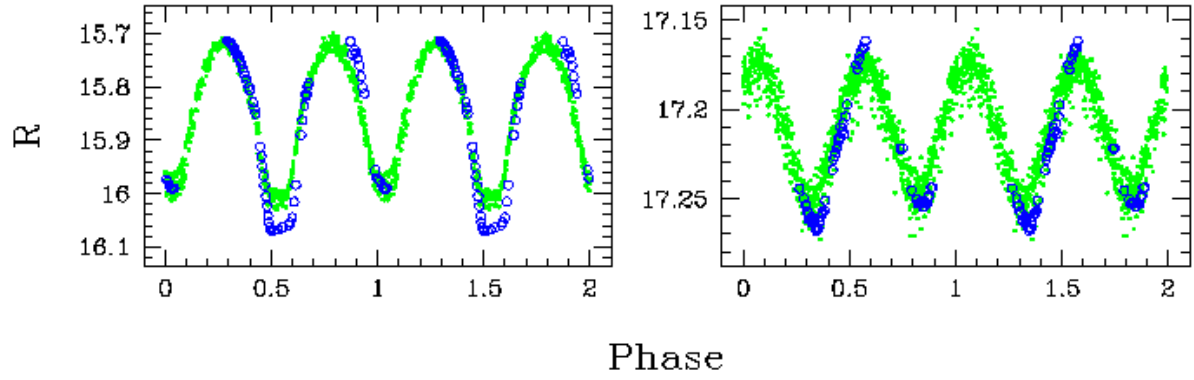}
\caption{Comparison between our light curves (open circles) and those from PISCES (closed circles) for a few known variable stars: (left) V1 (right) V4. An arbitrary constant has been added to our light curves to provide alignment with the PISCES light curves.}
\label{oldvar}
\end{figure}

As a check on our photometry we compare our light curves for a few
known variables with those published by \citet{mochejs05} in
Fig.~\ref{oldvar}. It is clear from the light curves that our
photometry matches well.

While the short time coverage of our observations prevents us from
performing a systematic survey for variable stars, we have identified
8 new variables in the field of NGC 6791 and 5 suspected variables that show
evidence for low-amplitude variability. Table~\ref{newvar_tab} lists
the coordinates and basic photometric data for these variables, which
we identify as V115-V122 and SV1-SV5. These stars were selected for
their short period variability using the Schwarzenberg-Czerny
algorithm \citep{scz96} as implemented in a code due to \citet{devor05}.

\begin{figure}[p]
\epsscale{1}
\plotone{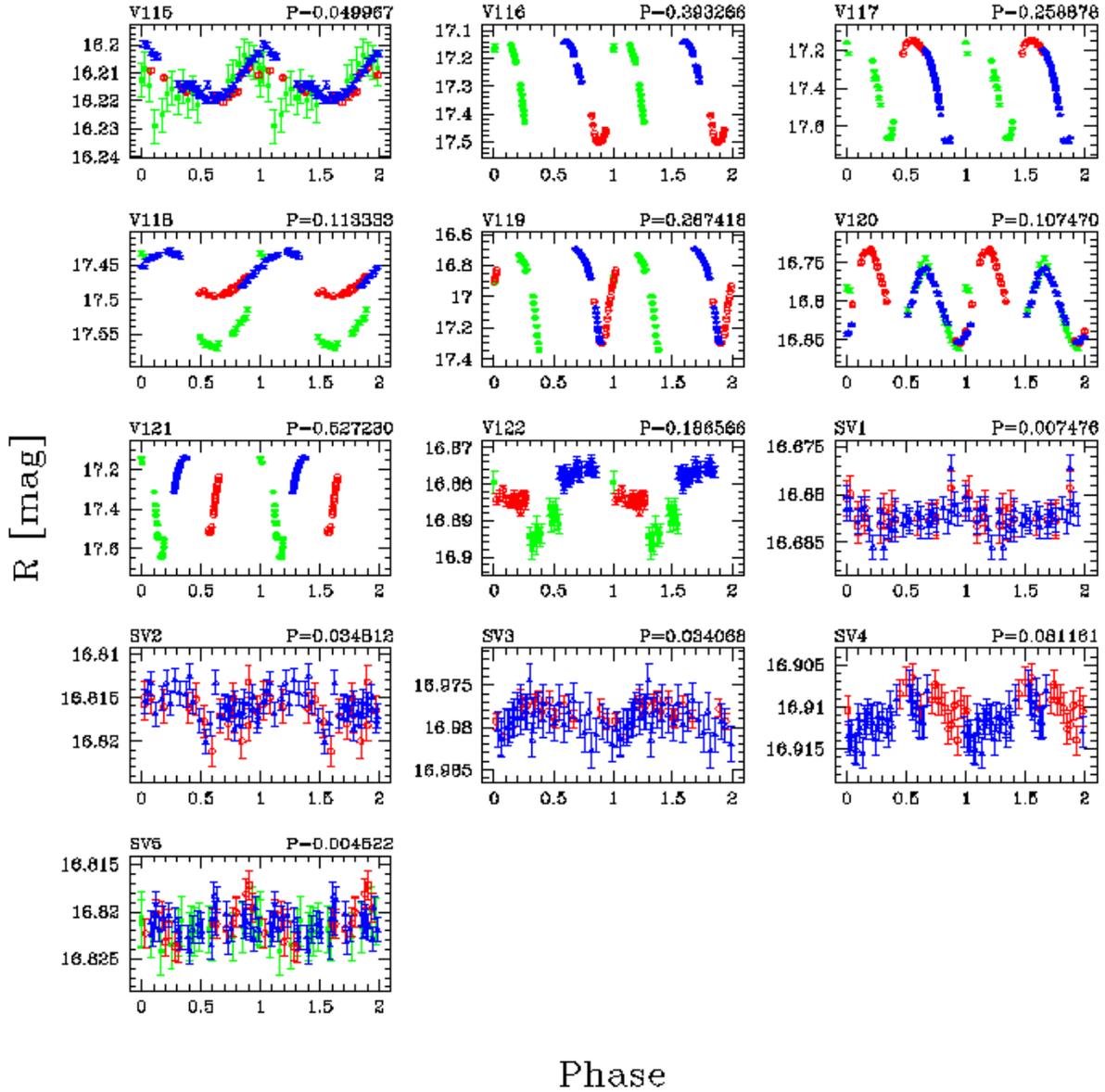}
\caption{Phased light curves for newly discovered variables. Data from the first night is shown as filled squares, the second night is shown as open circles, and the third night is shown as open triangles. $P$ is the period in days. We do not show data from the first night for the suspected variables when the noise from that night is greater than the amplitude of variability from the last two nights.}
\label{newvar}
\end{figure}

We present phased light curves for the newly discovered variables in
Fig.~\ref{newvar}\footnote{Light curves for all objects are available
upon request.}. For some of the suspected variable stars we have
omitted the data from the first night when the noise is greater than
the amplitude of variability in the second two nights. Due to a lack
of time coverage the periods are tentative.

From the light curves we classify V116, V117, and V119-V122 as likely
W UMa type contact binary systems. The variable V115 has a period and
light curve shape that is typical of a $\delta$-Scuti type pulsating
star. Lacking color information for this star, we cannot verify
whether this identification is correct. We note that differences
between the variations on the three days shows possible evidence for
multiple modes of pulsation. Also note the precision of this 20 mmag
amplitude light curve, particularly for the second and third
nights. The light curve of V118 is similar to that of an RS CVn type
spotted star. All of these variables lie outside of the field studied
by PISCES which is why they were not detected by that project.

The suspected variables (SV1-SV5) show very small full-amplitude
variations, as low as 3 mmag in the case of SV1. Of these, only SV2
lies within the field studied by PISCES. If real, the detection of
these subtle variations represents an exciting demonstration of the
potential of MMT/Megacam for precision time-series campaigns. At
present, with a limited amount of data, we cannot rule out the
possibility that some of these variations are the result of trends in
the data rather than actual stellar variability as we have made no attempt to correct for color-dependent extinction or correlations with pixel-position.

\section{Discussion}

We have successfully demonstrated the capability of the MMT/Megacam to
achieve very high precision photometry, as low as 360 $\mu$mag at the
bright end, for a large number of stars. In the process we have
discovered 8 new variable stars, and have identified 5 possible
variable stars with amplitudes as low as 3 mmag. While we have not
broken the record so to speak for the best precision per exposure
obtained from the ground, we have achieved sub-millimagnitude
photometry from the ground for more stars at once than has ever been
reported. Our results with aperture photometry on the subtracted
images shows that there are no barriers to our ability to achieve
precisions of a few hundred micromagnitudes with
this telescope and instrument.

While the detectection of solar-like, p-mode oscillations in other
stars is difficult to do in a reasonable amount of time, even with sub-millimagnitude photometry (see
for example \citealt{gilliland93}), we can still probe a relatively
unexplored regime of stellar variability.

\begin{figure}[p]
\epsscale{1}
\plotone{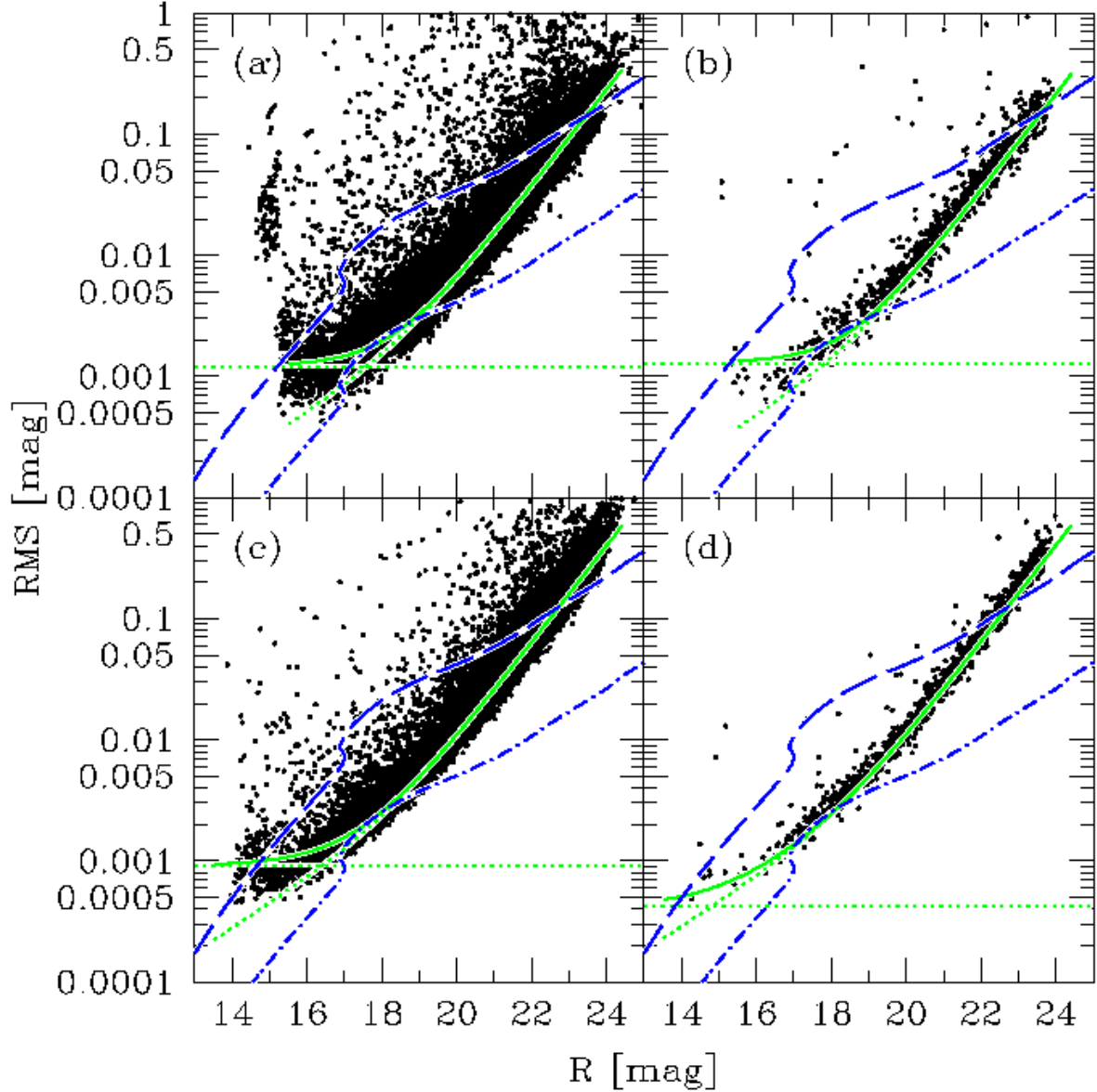}
\caption{The light curve RMS versus magnitude is plotted for (a) chips 1-33 on night 2, (b) chip 21 on night 2, (c) chips 1-33 on night 3, and (d) chip 21 on night 3. The solid line shows a fit to the observed errors including photon noise and a constant error term (dotted lines), the dashed line shows the 6.5$\sigma$ detection limit for a transiting Jupiter-sized planet, while the dot-dashed line shows the same limit for a transiting Neptune-sized planet. The generation of these lines is described in \S 4 and we discuss the implications in \S 6.}
\label{lcstat_separate}
\end{figure}

\begin{figure}[p]
\epsscale{1}
\plotone{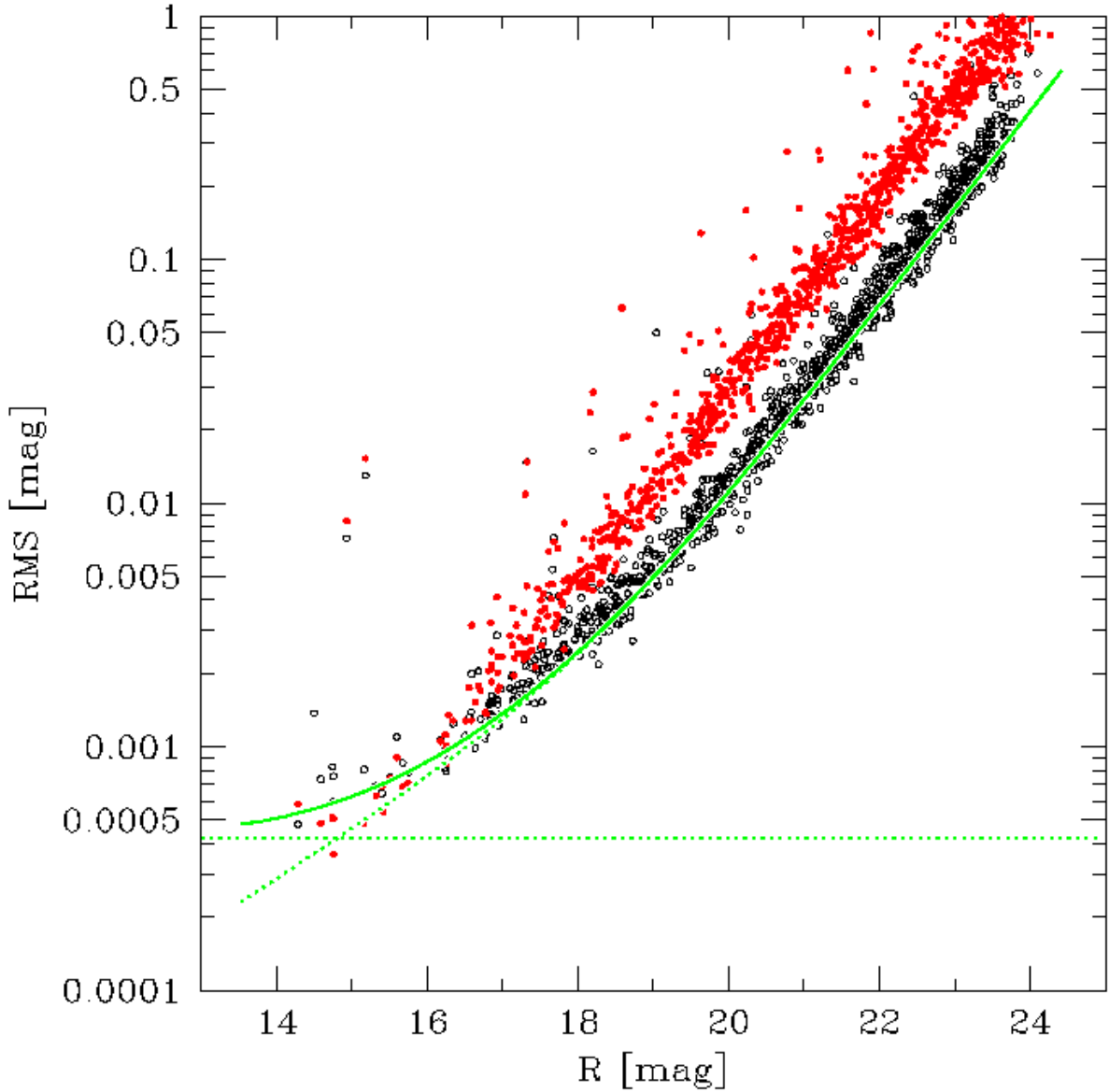}
\caption{Comparison between the precision from performing aperture photometry on the subtracted images (filled points) and PSF photometry on the subtracted images (open circles) for chip 21. The curves are the same as those in Fig.~\ref{lcstat_separate} (c). Note that aperture photometry apears to provide better precision for the brightest stars, while it provides worse photometry for fainter stars as the result of an effectively higher sky flux through the unweighted aperture.}
\label{apcompare}
\end{figure}

Another exciting application of this technology could be to survey
stellar clusters for planets as small as Neptune. In Figures
\ref{lcstat_separate}-\ref{apcompare} we showed the 6.5$\sigma$
detection limits for planets as small as Neptune assuming one observed
3 full transits with MMT/Megacam. For the second night, with the two
minute exposure time, there are 23,062 stars below the Jupiter
detection limit, and 1664 stars below the Neptune detection limit. On
the third night, with the 1 minute exposures, there are 19,843 stars
below the Jupiter detection limit, and 648 stars below the Neptune
detection limit. We have made no attempt here to distinguish between
field stars and cluster members.

The ability to detect Jupiters in this system is not limited by
precision but rather the time baseline over which observations are
carried out. If one observes long enough to have a resonable chance of
detecting 2-3 full transits, then one would be able to find
essentially every short-period, transiting, Jupiter-sized planet in
the stellar cluster. Since there are only three known planets with a
lower mass-limit near that of Neptune
\citep{santos04a,mcarthur04,butler04}, essentially nothing is known
about the statistics of Neptune-sized planets. Moreover, because all
of these planets have been detected only via their influence on the
radial velocities of their host stars, we do not know anything about
their radii. Therefore, the very fact that there are hundreds of stars
around which we could detect transiting, Neptune-sized planets if they
exist represents an exciting new opportunity for the study of
extra-solar planets.

As noted by \citet{pepper05}, as a result of the relations betwen
mass, luminosity, and radius for main sequence stars, if one can
identify a transiting planet around any cluster member with
source-limited precision, then one could find that same transiting
planet around essentially all cluster members with source-limited
precision. The effect of this is that it is not essential, for planet
finding, to achieve source-limited photometry at the brightest end
where there are few stars, rather it is most important to achieve
source-limited photometry for a large number of stars. Therefore, even
if our photometry shows some small ($<$1 mmag) constant error term at
the bright end, we would still have sensitivity to Neptune-sized
planets around many stars. Thus it is not fundamentally the ability to
do high-precision photometry just below saturation that opens up the
possibility of finding small planets, but rather it is the fact that
we are using a large telescope that can collect a greater signal per
exposure for every star compared with using a smaller telescope.

\begin{figure}[p]
\epsscale{1}
\plotone{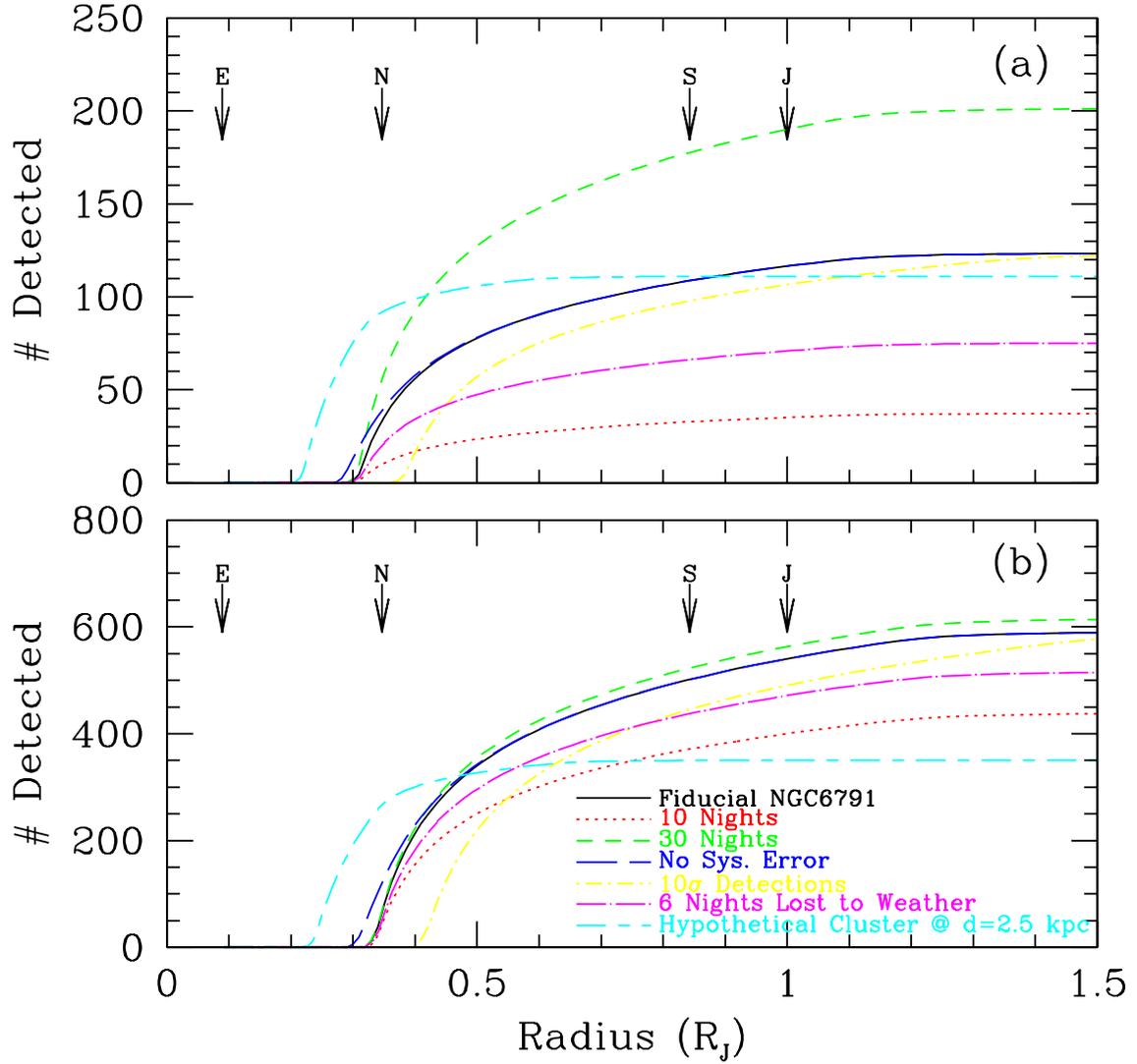}
\caption{The expected number of hot planets ($P=3-9$ days, a), and very hot planets ($P=1-3$ days, b) detected as a function of planetary radius assuming all stars have planets of the given radius within the stated periods. For the fiducial calculation we assume 20 nights of perfect weather, 7 hours per night, and a detection threshold of $6.5\sigma$.}
\label{gaudi}
\end{figure}

We can estimate the number of planets one could detect in an
ambitious, many-night survey of NGC 6791 using Megacam on the
MMT. Adopting the parameters of NGC 6791 listed previously
[$E(B-V)=0.1$, distance$=4.8$kpc, age$=8$Gyr], and assuming a mass
function slope and normalization that reproduces the empirical I-band
luminosity function of \citet{kaluzny92}, we calculate the number of
planets one would detect as a function of the planetary radius using
the formalism of \citet{pepper05}. We assume that the planets are
uniformly distributed in log period, and we consider planets with
periods $P=1-3$ days and $3-9$ days separately. For our fiducial
calculation, we assume a detection threshold of $S/N>6.5$, 7 hours
per night, 0.1\% systematic error, and perfect weather. We then
consider the effect of changing each of these fiducial assumptions on
the number of detected planets. Fig.~\ref{gaudi} shows the number of
detected planets as a function of radius, under the assumption that
every star has a planet of a given radius. As mentioned above, NGC
6791 has a super-solar metallicity of [Fe/H]$=0.3-0.4$ dex, which
implies a frequency of hot Jupiters ($P=3-9$ days) of $\sim$ 4\%
\citep{fischer05, santos04b, mochejs05}, and a frequency of very hot
Jupiters ($P=1-3$ days) of $\sim$ 0.6\% \citep{gaudi05}, assuming the
population of planets is similar to the local solar neighborhood. For
our fiducial assumptions, these frequencies yield $\sim$ 5 expected
detections of hot Jupiters and $\sim$ 3 expected detections of very hot
Jupiters.

One is unlikely to detect a significant number of hot Neptunes in NGC
6791, unless they are considerably more common than their massive
counterparts. However, NGC 6791 is not necessarily ideal for the
detection of Hot Neptunes, and closer clusters will likely yield
improved expected detection rates. To demonstrate this, in
Fig.~\ref{gaudi} we also show the number of expected detections for a
hypothetical cluster with the same parameters as NGC 6791, but with
distance$=2.5$kpc, 2500 stars, 10 hours per night, and 0.5\%
systematic error. In this case, one would expect to detect $\sim 90f$
hot Neptunes, and $\sim 270f$ very hot Neptunes, assuming a fraction $f$
of stars have Neptune-sized planets in the given range of
periods. Finally we mention that the weather in Arizona also makes NGC
6791 an unideal target for the MMT. At $19^{\rm h}21^{\rm m}$ of RA, the
cluster is best observed in July/August and is thus subject to the
Arizona monsoon season. We have also obtained preliminary data for the open clusters M35 and NGC 2158. The results from these clusters will be presented in a future contribution.

\acknowledgements{We gratefully acknowledge B.~Mochejska for the use of data from PISCES in calibrating our photometry, and for helpful discussions. We also acknowledge G.~Pojmanski for his
excellent ``lc'' program, and W.~Pych for his ``fwhm'' program. JDH is
funded by a National Science Foundation Graduate Student Research
Fellowship. KZS acknowledges support from the William F. Milton
Fund. BSG is supported by a Menzel Fellowship from the Harvard
College Observatory.}

\begin{deluxetable}{lrrrrr}
\tabletypesize{\footnotesize}
\tablewidth{0pc}
\tablecaption{New variable stars identified in the field of NGC 6791. Stars listed as ``SV'' are suspected variables.}
\tablehead{\colhead{ID} & \colhead{$\alpha_{2000}$} & \colhead{$\delta_{2000}$} & \colhead{Period (days)} & \colhead{R} & \colhead{$A_{R}$}}
\startdata
V115 &  19$^{h}$20$^{m}$10\fs22 &  37\degr43\arcmin12\farcs2 &  0.049967 &  16.21 &  0.018 \\
V116 &  19$^{h}$20$^{m}$36\fs36 &  37\degr39\arcmin56\farcs7 &  0.393266 &  17.26 &  0.357 \\
V117 &  19$^{h}$20$^{m}$51\fs00 &  37\degr39\arcmin03\farcs9 &  0.258878 &  17.36 &  0.524 \\
V118 &  19$^{h}$21$^{m}$07\fs07 &  37\degr54\arcmin58\farcs9 &  0.113333 &  17.48 &  0.135 \\
V129 &  19$^{h}$21$^{m}$29\fs03 &  37\degr55\arcmin56\farcs3 &  0.267418 &  16.92 &  0.594 \\
V120 &  19$^{h}$21$^{m}$43\fs46 &  37\degr39\arcmin56\farcs9 &  0.107470 &  16.80 &  0.116 \\
V121 &  19$^{h}$21$^{m}$54\fs52 &  37\degr40\arcmin53\farcs4 &  0.527230 &  17.30 &  0.498 \\
V122 &  19$^{h}$21$^{m}$56\fs94 &  37\degr49\arcmin23\farcs7 &  0.186566 &  16.88 &  0.020 \\
SV1  &  19$^{h}$20$^{m}$50\fs28 &  37\degr40\arcmin06\farcs9 &  0.007476 &  16.68 &  0.003 \\
SV2  &  19$^{h}$21$^{m}$10\fs35 &  37\degr44\arcmin40\farcs0 &  0.034812 &  16.82 &  0.005 \\
SV3  &  19$^{h}$21$^{m}$39\fs05 &  37\degr58\arcmin04\farcs6 &  0.034068 &  16.98 &  0.005 \\
SV4  &  19$^{h}$21$^{m}$47\fs01 &  37\degr45\arcmin45\farcs3 &  0.081161 &  16.91 &  0.007 \\
SV5  &  19$^{h}$21$^{m}$53\fs45 &  37\degr49\arcmin45\farcs0 &  0.004522 &  16.82 &  0.005 \\
\enddata
\tablecomments{ Coordinates are from {\it 2MASS} where available. The first 8 entries are confirmed variables while the last five are low amplitude suspected variables. The periods listed are those used to phase the light curves in Fig.~\ref{newvar} and should not be treated as a constrained value for the period. Amplitudes are defined as the difference between the third brightest and third faintest observations in the data presented in Fig.~\ref{newvar}, while average magnitudes are flux-weighted.}
\label{newvar_tab}
\end{deluxetable} 

\end{document}